\documentclass[superscriptaddress,aps,prb,twocolumn]{revtex4-2}

\usepackage{amsmath,amssymb,graphicx,dcolumn,bm,hyperref,xcolor}
\usepackage{appendix}

\begin{document}

\title{Characterizing Spin-Orbit Torques by Tensorial Spin Hall Magnetoresistance}

\author{Hantao Zhang}
\affiliation{Department of Electrical and Computer Engineering, University of California, Riverside, CA 92521, United States}	
\author{Ran Cheng}
\email{ran.cheng@ucr.edu}
\affiliation{Department of Electrical and Computer Engineering, University of California, Riverside, CA 92521, United States}
\affiliation{Department of Physics and Astronomy, University of California, Riverside, CA 92521, United States}
\affiliation{Department of Materials Science and Engineering, University of California, Riverside, CA 92521, United States}

\begin{abstract}
Magnetoresistance (MR) provides a crucial tool for experimentally studying spin torques. While MR is well established in the device geometry of the spin Hall effect (SHE), as exemplified by the magnet/heavy-metal heterostructures, its role and manifestation beyond the SHE paradigm remain elusive. We propose a hitherto unknown form of MR where the underlying charge-to-spin conversion and its inverse process violate the simple geometry of the SHE, calling for tensorial descriptions. This MR can generate a series of unique harmonic responses essential for the experimental characterization of unconventional spin-orbit torques in non-SHE materials. We demonstrate these harmonic signals with semimetal WTe$_2$ in mind but the results are not restricted to specific materials.
\end{abstract}

\maketitle

Spin torques are essential for achieving effective control of magnetism through electrical currents, which holds great potential in modern electronics~\cite{SOTroadmap,ryu2020current,hirohata2020review,ramaswamy2018recent}. An important way of generating spin torques is to take advantage of the spin-orbit coupling in non-magnetic materials (such as heavy metals) that enables the spin Hall effect (SHE)~\cite{SHEhirsch,SHErmp,hoffmann2013spin}, converting a charge current into a transverse spin current. The spin current will then deliver spin angular momenta to, hence driving, the magnetic order of an adjacent material in the form of spin-orbit torque (SOT)~\cite{SOTrmp,SOTsongreview,amin2016spin}. As a salient feature of the SHE~\cite{SHGeometry}, the spin polarization, the charge current, and the transport direction of spins are mutually orthogonal.

This seemingly established relation, however, has been seriously challenged by recent experimental discoveries, where the non-equilibrium spin polarization induced by the spin-orbit interactions can significantly deviate from the SHE geometry owing to the underlying crystal symmetry~\cite{macneill2017control,kao2022deterministic,kao2025unconventional,shin2022spin,meng2025non}. For example, in a semimetal WTe$_2$ thin film, a current applied along the $a$ axis can generate a substantial out-of-plane spin polarization exerting unconventional SOT on an adjacent magnet, whereas a current applied along the $b$ axis only produces an in-plane spin component following the SHE~\cite{macneill2017control,kao2022deterministic,kao2025unconventional}. Significant out-of-plane spin generations violating the ordinary SHE have also been claimed in non-collinear antiferromagnets such as Mn$_3$Sn~\cite{kimata2019magnetic,kondou2021giant,hazra2023generation,hu2022efficient,meng2024field,Mn3SnExp} and IrMn$_3$~\cite{IrMn3Exp1,IrMn3Exp2}. While such SOT has showcased its unique advantage in achieving field-free switching of perpendicular magnets, a systematic understanding of its physical characteristics remains superficial, largely ascribing to the elusive characterization schemes within available experimental setup.

Magnetoresistance (MR) refers to the electrical resistance depending on the magnetic order, which is a pivotal quantity to characterize SOT in real experiments such as the spin-torque ferromagnetic resonance~\cite{Liu2011STFMR,Liu2012Science,mellnik2014spin} and harmonic Hall measurements~\cite{hayashi2014quantitative,vlietstra2014simultaneous,avci2014interplay,macneill2017thickness,chen2018first,schippers2020large,Kent2022,cheng2022third,ZhangAFMharmonic}. Some of these experiments utilize a ferromagnetic metal hosting anisotropic MR, where an electrical current in an adjacent material creates spin torques that acts on the magnetization and thus affecting the anisotropic MR. In turn, the variation of MR enables the quantification of the spin torques. An alternative setup involves a ferromagnetic insulator (FI) coupled to a heavy metal~\cite{SMRNakayama,SMRHahn,SMRAlthammer}, which avoids the shunting current. The SHE in the heavy metal not just generates the SOT but simultaneously detects the magnetization dynamics through the spin-Hall MR (SMR)~\cite{SMR} accompanying the SOT. However, if the SOT arises from unconventional mechanisms that violates the SHE geometry, the corresponding MR is not clear at all, let alone using the MR to characterize the SOT. For instance, in a WTe$_2$/FI bilayer~\cite{kao2025unconventional}, the non-equilibrium spin polarization driven by an in-plane current bears a significant out-of-plane component, thus not perpendicular to the flowing direction of the spins, resulting in a complicated geometry that must be described by tensorial relations.

In this paper, we first establish a phenomenological description of the charge-to-spin conversion and its reciprocal process beyond the SHE scenario, allowing the non-equilibrium spin polarization to be oriented arbitrarily with respect to the charge current, which brings about the \textit{tensorial} SHE (\textit{t}-SHE). We then derive a new form of MR arising from the combined action of the \textit{t}-SHE and its inverse effect in the presence of spin diffusion and backflow. This generalized effect, termed \textit{tensorial} SMR (\textit{t}-SMR), reduces to the ordinary SMR when the SHE geometry is imposed. The first-harmonic signal due to the \textit{t}-SMR can be utilized to determine the direction of the non-equilibrium spin polarization. Next, by solving the SOT-driven magnetization dynamics and the higher harmonics from the \textit{t}-SMR, we obtain a series of electric signals detectable through magnetic resonances and harmonic-Hall experiments, which enables practical methods to experimentally characterize the SOTs originating from the \textit{t}-SHE. These features are compared with conventional theories associated with the SHE geometry. Finally, we apply our theory numerically to a WTe$_2$/FI bilayer structure, which is representative of non-SHE systems. While the examples are presented with the geometry of WTe$_2$ in mind, our theoretical results are general and cover broad scenarios beyond the SHE.

\begin{figure}[t]
  \centering
  \includegraphics[width=\linewidth]{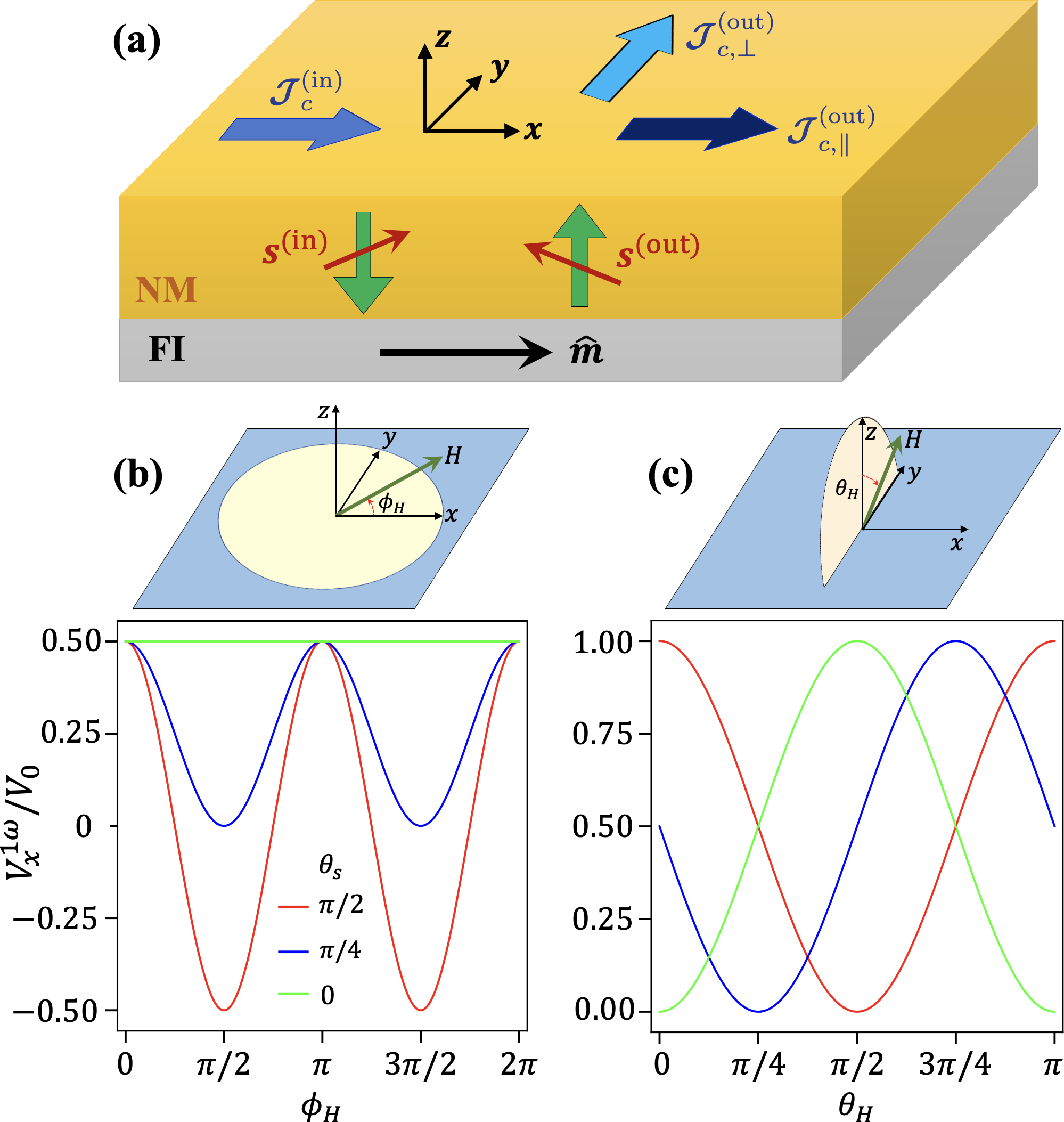}
  \caption{(a) Illustration of the $t$-SHE and its reciprocal effect. A charge current (density) $\bm{\mathcal{J}}_{c}^{(\rm in)}$ injected along $x$ generates a spin current flowing in $z$, which carries a spin polarization $\bm{s}^{(\rm in)}$ due to the $\textit{t}$-SHE. The reflected spin current carrying $\bm{s}^{(\rm out)}$ converts back into charge currents through the inverse $\textit{t}$-SHE, which produces $\bm{\mathcal{J}}_{c,\parallel}^{(\rm out)}$ and $\bm{\mathcal{J}}_{c,\perp}^{(\rm out)}$, affecting the MR. (b) and (c) plot the longitudinal first-harmonic signal $V_{x}^{1\omega}$ scaled by the non-harmonic background $V_0$ in WTe$_2$-like materials for the $xy$ and $yz$ field scans, respectively, where $\theta_{H}$ and $\phi_{H}$ specify the magnetic field direction while $\theta_{s}$ parametrizes the direction of non-equilibrium $\bm{s}$. The values of the vertical axes are measured in $2t\eta^2\tilde{g}_R$, typically of order $10^{-3}\sim10^{-4}$.}
\label{fig:geometry}
\end{figure}

\textit{Tensorial Spin Hall Magnetoresistance.}---We first draw an intuitive physical picture without resorting to rigorous math. As schematically illustrated in Fig.~\ref{fig:geometry}(a), a bilayer heterostructure composed of a non-magnetic metal (NM) and a FI is manipulated by a charge current (density) $\bm{\mathcal{J}}_{c}^{(\rm in)}$ applied along the $x$ axis. Through spin-orbit interactions, $\bm{\mathcal{J}}_{c}^{(\rm in)}$ can generate spin currents flowing in all directions, among which the $z$-directed component carrying a non-equilibrium spin polarization $\bm{s}^{(\rm in)}$ dominates the transport properties in such a thin-film geometry. If the charge-to-spin conversion strictly follows the SHE, $\bm{s}^{(\rm in)}$ would be simply along $y$, but here we intentionally allow it to defy the SHE geometry. Thanks to the spin diffusion effect, a steady spin accumulation will be induced on the NM/FI interface~\cite{maekawa2017spin}, exerting SOTs on the (unit) magnetization vector $\hat{\bm{m}}$. In turn, the SOT-driven magnetization gives rise to a reflected spin current carrying a spin polarization $\bm{s}^{(\rm out)}$, which can be converted back into charge currents on top of $\bm{\mathcal{J}}_{c}^{(\rm in)}$. Therefore, the outgoing current contains a longitudinal component $\bm{\mathcal{J}}_{c,\parallel}^{(\rm out)}$ (along $x$) and a transverse component $\bm{\mathcal{J}}_{c,\perp}^{(\rm out)}$ (along $y$). The Onsager reciprocal relations require that $\bm{\mathcal{J}}_{c,\parallel}^{(\rm out)}$ stems from the projection of $\bm{s}^{(\rm out)}$ on the direction of $\bm{s}^{(\rm in)}$ via the spin-to-charge conversion, whereas $\bm{\mathcal{J}}_{c,\perp}^{(\rm out)}$ can be traced back to a spin polarization $\bm{s}'$ noncollinear with $\bm{s}^{\rm(in)}$ hauled by the reflecting spin current; $\bm{s}'$ becomes perpendicular to $\bm{s}^{(\rm in)}$ only in the SHE limit. Regarding the Onsager relations, one can also treat $\bm{s}'$ as the spin polarization of a $z$-directed incident spin current had the driving charge current $\bm{\mathcal{J}}_{c}^{(\rm in)}$ be applied along $y$ instead of $x$. In a semimetal WTe$_{2}$ thin film, for instance, $\bm{s}^{(\rm in)}$ involves both in-plane and out-of-plane components depending on the direction of $\bm{\mathcal{J}}_{c}^{(\rm in)}$ relative to the crystal axis~\cite{macneill2017control,kao2022deterministic,kao2025unconventional}. Since $\hat{\bm{m}}$ directly affects $\bm{s}^{(\rm out)}$, hence determining $\bm{\mathcal{J}}_{c,\parallel}^{(\rm out)}$ and $\bm{\mathcal{J}}_{c,\perp}^{(\rm out)}$, the system's overall electric responses will manifest as a non-trivial MR~\cite{commentAMR}.

Guided by the above physical picture, we now establish a general relation for the coupled transport of spin and charge in the NM. Let $\mu_c$ be the electrochemical potential and $\mu_i$ be the $i$ component of the spin-chemical potential (with $i=x,y,z$). By scaling the spin and charge current densities into the same unit (A/m$^2$), we can incorporate and unify them into a $3\times4$ matrix $\mathcal{J}_{ij}$, where $i=x,y,z$ specifies the current flow direction and $j=x,y,z$ specifies the spin polarization; whereas the $4$-th column $\mathcal{J}_{ic}$ refers to the charge current density in the $i$ direction. Accordingly, the \textit{t}-SHE and its inverse effect should be described by the following equations:
\begin{subequations}
\label{eq:current_potential}
\begin{align}
    & \mathcal{J}_{ij} = -\frac{\sigma}{e}\left[\chi_{ijk}(\partial_k\mu_c)+\frac12\partial_i\mu_j\right],
    \label{eq:spin_current_potential} \\
    & \mathcal{J}_{ic} = -\frac{\sigma}{e}\left[\partial_i\mu_c+\frac12\chi'_{ijk}(\partial_j\mu_k)\right],
    \label{eq:charge_current_potential}
\end{align}
\end{subequations}
where $\sigma$ is the intrinsic electrical conductivity, $e$ is the electron charge, and $\chi_{ijk}$ and $\chi'_{ijk}$ are the conversion tensors between the spin and charge degrees of freedom. The Einstein summation rule is assumed throughout this paper. In the SHE limit, $\chi_{ijk}=-\chi'_{ijk}=\vartheta_{\rm SH} \epsilon_{ijk}$ with $\vartheta_{\rm SH}$ being the spin Hall angle and $\epsilon_{ijk}$ the Levi-Civita symbol (total antisymmetric tensor). In a general context without the SHE constraints, the Onsager relations only require that $\chi_{ijk} = -\chi'_{kij}$. Although a tensorial description is employed for the interconversion between spin and charge degrees of freedom, we must emphasize that the underlying physics here is entirely different from the spin-tensor Hall effect~\cite{SuPRB2023}, which involves larger-spin carriers rather than electron spins, as well as that associated with the pseudo-spin degree of freedom~\cite{fu2020intrinsic}.

The spin diffusion process in the NM is governed by $\nabla^2\mu_j=\mu_j/\lambda^2$, where $\lambda$ is the spin diffusion length. In the thin-film geometry, $\nabla^2\approx\partial_z^2$, while the charge current flows in the $xy$ plane. Consequently, only $6$ of the $27$ elements in $\chi_{ijk}$ are non-zero, reducing the conversion tensor into two independent vectors:
\begin{align}
  \chi_{zjx} = s_{j},\quad \chi_{zjy} = s_{j}', \label{eq:ssprime}
\end{align}
which can be regarded as the $j$-th components of $\bm{s}^{(\rm in)}$ and $\bm{s}'$ described in the intuitive picture. As detailed in the Supporting Information (SI), we can solve $\mu_j(z)$ as a function of $z$ using the same boundary conditions as the SMR~\cite{SMR,Feedback}: $e\mathcal{J}_{zj}=g_R\epsilon_{jkl}\epsilon_{lab}\hat{m}_k\hat{m}_a\mu_b+g_I\epsilon_{jkl}\hat{m}_k\mu_l$ at $z=0$ and $\mathcal{J}_{zj}=0$ at $z=t$ (the NM thickness), where $g_R$ and $g_I$ are the real and imaginary parts of the interfacial spin-mixing conductance per unit area $g_{\uparrow\downarrow}$. The effective MR can be read off by inserting $\mu_j$ into Eqs.~\eqref{eq:current_potential}. To simplify the notation, the MR can be conveniently expressed in terms of conductivity, and one should be borne in mind that $\rho_{xx}=\sigma_{xx}/(\sigma_{xx}^2+\sigma_{xy}^2)$ and $\rho_{xy}=\sigma_{xy}/(\sigma_{xx}^2+\sigma_{xy}^2)$. The longitudinal and transverse components are
\begin{subequations}
\label{eq:conductivities}
\begin{align} 
    \frac{\sigma_{xx}}{\sigma} &= 1 + 2\eta|\bm{s}|^{2}  - 2 t \eta^{2} \tilde{g}_R (\hat{\bm{m}} \times \bm{s})^{2},  \label{eq:sigma_xx} \\
    \frac{\sigma_{xy}}{\sigma} &= 2\eta \bm{s} \cdot \bm{s}' - 2 t\eta^{2}   \left[ \tilde{g}_R (\hat{\bm{m}} \times \bm{s})\cdot(\hat{\bm{m}} \times \bm{s}') \right. \nonumber \\
    &\qquad\qquad\qquad\qquad\left. - \tilde{g}_I \hat{\bm{m}}\cdot(\bm{s}\times\bm{s}') \right], \label{eq:sigma_xy}    
\end{align}
\end{subequations}
where $\tilde{g}_{R} = \mathrm{Re}\{g_{\uparrow\downarrow} / [\sigma + 2\lambda g_{\uparrow \downarrow} \coth(t/\lambda)]\}$ (and $\tilde{g}_I$ takes a similar form, with $\rm{Re}\rightarrow\rm{Im}$) is the renormalized spin-mixing conductance per unit area incorporating the spin backflow~\cite{Backflow}, and $\eta = (\lambda / t) \tanh (t/2\lambda)$ is a dimensionless quantity that decreases with an increasing ratio $t/\lambda$. In the SHE limit, $\bm{s} = \pm\vartheta_s\hat{\bm{y}}$ and $\bm{s}' = \mp \vartheta_s\hat{\bm{x}}$, reducing Eqs.~\ref{eq:sigma_xx} and~\ref{eq:sigma_xy} to the ordinary SMR~\cite{SMR}.

Two remarks are in order. First, although the conversion relations between spin and charge degrees of freedom are generalized to tensors, $\chi_{ijk}$ and $\chi'_{ijk}$, reflecting the anisotropic nature of spin and charge transport in the non-magnetic metal, we assume that the spin transport at the NM/FI interface remains isotropic. Consequently, the spin mixing conductance $g_{\uparrow\downarrow}$ is treated as a scalar rather than a tensor, consistent with previous studies that did not consider anisotropic spin and charge transport~\cite{SpinPumping,TserkovnyakRMP2005,AFMSpinPumping2014}. This assumption enables us to focus on the unconventional MR arising solely from $\chi_{ijk}$ and $\chi'_{ijk}$, without additional complications from interfacial effects. Nevertheless, anisotropic bulk spin and charge transport can, in principle, induce anisotropy in the interfacial $g_{\uparrow\downarrow}$. A complete treatment of such tensorial spin mixing conductance and its influence on MR goes beyond the scope of this work. Second, the \textit{t}-SMR expressions Eqs.~\eqref{eq:conductivities} are general, not relying on any particular relation between $\tilde{g}_{R}$ and $\tilde{g}_{I}$. Such generality is especially important given growing evidence that $\tilde{g}_{I}$ can play a significant role in various magnetic insulator/heavy-metal heterostructures~\cite{Kent2022,cheng2022third,Uchimura2025PRL}.

\textit{Harmonic Signal Expansion.}--Under an AC perturbation, the \textit{t}-SMR obtained in Eqs.~\eqref{eq:conductivities} will generate a series of harmonic signals, which can be exploited to characterize the SOT originating from the \textit{t}-SHE. Specifically, the SOT-induced magnetization dynamics responds to an AC current density $\mathcal{J}_j^{\rm(in)}\cos\omega t$ in the form $\bm{m}(t)=\bm{m}_0+\delta\bm{m}\cos(\omega t+\varphi)$, where $\bm{m}_0$ denotes the equilibrium magnetization and $\varphi$ depends on the frequency $\omega$, rendering the MR tensor $\rho_{ij}[\bm{m}(t)]$ a periodic function of time. The output voltage $V_i(t)=\rho_{ij}[\bm{m}(t)]\mathcal{J}_j^{\rm(in)}\cos\omega t$ can be expanded into a group of harmonics~\cite{ZhangAFMharmonic}
\begin{align}
 V_i(t)=&\ \overline{\rho}_{ij}\mathcal{J}_j^{\rm(in)}\cos\omega t + \frac12\frac{\partial\overline{\rho}_{ij}}{\partial\bm{m}_0}\cdot\delta\bm{m} \notag\\
&\qquad\times\mathcal{J}_j^{\rm(in)}\Big[\cos{\varphi}+\cos(2\omega t+\varphi)\Big]+\cdots,
\label{eq:harmonicexpansion}
\end{align}
where $\overline{\rho}_{ij}=\rho_{ij}[\bm{m}_0]$ and the second harmonic involves a DC term known as rectification. For a low-frequency drive such that $\omega$ is far below the magnetic resonance frequency, $\bm{m}(t)$ will follow adiabatically and remain in the quasi-equilibrium state determined by the balance between the magnetic interactions and the instantaneous SOT, thus generating the harmonic voltages on top of the non-harmonic background $V_0$. For a high-frequency drive, the rectification signal can directly monitor the onset of SOT-induced ferromagnetic resonance (FMR). Therefore, by detecting different harmonics and controlling $\bm{m}_0$ via a magnetic field, we can make inquiry into various aspects of the SOT.

\textit{First Harmonic.}---With a low-frequency AC drive, the $t$-SMR can be directly read off from the first-harmonic signal $V_i^{1\omega}=\overline{\rho}_{ij}\mathcal{J}_j^{\rm(in)}$, which reveals critical information about $\bm{s}$ and $\bm{s}'$ defined in Eq.~\eqref{eq:ssprime}, setting the stage for further investigations. As a concrete example, we consider a semimetal WTe$_{2}$ thin film coupled to a FI, in which an applied current density $\mathcal{J}_x^{\rm(in)}$ can induce a non-equilibrium spin polarization $\bm{s}/s = \text{cos}\theta_{s} \hat{\bm{z}} + \text{sin}\theta_{s} \hat{\bm{y}}$ with $\theta_s$ depending on the crystal orientation relative to the $x$ axis. If $x$ coincides with the high-symmetry axis, $\theta_s$ would become $\pi/2$, conforming with the SHE limit. However, if $x$ is parallel to the low-symmetry axis, there has been no consensus on (in fact, no reliable way to quantify) $\theta_s$ to the best of our knowledge.

This conundrum can be resolved by measuring the longitudinal \textit{t}-SMR entailed in the first harmonic signal $V_{x}^{1\omega}$. To this end, we consider a strong magnetic field $\bm{H}$ that polarizes the magnetization so that $\hat{\bm{m}}$ will follow $\bm{H}$ as it rotates. In Fig.~\ref{fig:geometry}(b) and~(c), we plot $V^{1\omega}_{x}$ versus the field direction as $\bm{H}$ scans in the $xy$ and $yz$ planes, for a few different values of $\theta_s$. The $\theta_s=\pi/2$ case (red curves) corresponds to the SHE limit, which is fully consistent with known experiments~\cite{SMRNakayama,SMRHahn,SMRAlthammer}. The departure of $V_{x}^{1\omega}$ as a function of $\bm{H}$ from its SHE limit, if $\theta_s$ deviates from $\pi/2$, is reflected in both the magnitude of the $\phi_H$-dependence ($xy$ scan) shown in Fig.~\ref{fig:geometry}(b) and the phase shift of the $\theta_H$-dependence ($yz$ scan) shown in Fig.~\ref{fig:geometry}(c); these two quantities vary with $\theta_s$ in the forms of $\sin^2\theta_s$ and $\cos^2(\theta_H-\theta_s)$, respectively. Therefore, we can quantify $\theta_s$ for WTe$_2$ by benchmarking the first-harmonic signals for $\mathcal{J}_x^{\rm(in)}$ along the low-symmetry axis against those for $\mathcal{J}_x^{\rm(in)}$ along the high-symmetry axis.

\textit{Rectification.}---The DC rectification as a constituent of the second harmonic response has been widely employed to monitor and quantify the SOT-induced FMR. In the SHE paradigm, the rectification signal as a function of a sweeping magnetic field (that controls $\bm{m}_0$) can be decomposed into symmetric and antisymmetric parts with respect to the resonance point, which reflects the strength of the damping-like (DL) torque relative to the field-like (FL) torque (including the contribution of the Ørsted field)~\cite{Liu2011STFMR,Liu2012Science}. When the SHE relations are explicitly broken, however, we shall scrutinize the established results by revisiting the interplay between the magnetization dynamics and the coupled spin-charge transport described by Eqs.~\eqref{eq:current_potential}.

Consider a soft magnet with $\hat{\bm{m}}$ polarized by a magnetic field such that the Zeeman interaction dominates the magnetic anisotropy and the demagnetization field. The Landau-Lifshitz-Gilbert (LLG) equation is
\begin{align}
    \frac{\partial\hat{\bm{m}}}{\partial t} &= \bm{H}\times \hat{\bm{m}} + \alpha\hat{\bm{m}} \times \frac{\partial\hat{\bm{m}} }{\partial t} + \hat{\bm{m}} \times (\bm{H}_{\rm DL} \times \hat{\bm{m}} ) \nonumber \\
   &\qquad +\bm{H}_{\rm F}\times\hat{\bm{m}} + g_{R}'\hat{\bm{m}} \times \frac{\partial\hat{\bm{m}} }{\partial t} + g_{I}'\frac{\partial\hat{\bm{m}} }{\partial t}, \label{eq:LLG_main}
\end{align}
where the Zeeman field $\bm{H}$ absorbs the gyro-magnetic ratio $\gamma$, $\alpha$ is the dimensionless Gilbert damping constant, and $g_{R,I}'=g_{R,I}\gamma\hbar^2/(2e^2M_st)$ are two dimensionless coefficients (with $M_s$ being the saturation magnetization) determining the strength of spin pumping---a reciprocal effect of the SOT~\cite{SpinPumping,kajiwara2010transmission}. The DL torque exerted by $\bm{s}$ is determined by $\bm{H}_{\rm DL}=H_{\rm DL}\hat{\bm{s}}$, while $\bm{H}_{\rm F}=H_{\rm Oe}\hat{\bm{y}}+H_{\rm FL}\hat{\bm{s}}$ incorporates the FL torque and the Ørsted-field effect. Along with $\bm{H}$, these fields are both scaled into the frequency dimension. By solving Eq.~\eqref{eq:LLG_main} (elaborated in the SI), we obtain the rectification voltage as
\begin{align}
 \frac{V_{\rm rec}}{V_0} = \mathcal{S}\frac{\Delta}{(H-\omega_{0})^{2}+\Delta^{2}} + \mathcal{A}\frac{H-\omega_0}{(H-\omega_{0})^{2}+\Delta^{2}}, \label{eq:ST_FMR_V}
\end{align}
where $\omega_{0} = \omega \sqrt{(1-g_{I}')^{2} - (\alpha + g_{R}')^{2}}$ and $\Delta = \omega (1-g_{I}')(\alpha + g_{R}') / \sqrt{(1-g_{I}')^{2} - (\alpha + g_{R}')^{2}}$. The right-hand side of Eq.~\eqref{eq:ST_FMR_V} consists of a symmetric function and an antisymmetric function of $H$ with respect to the resonance field $H_{\rm res}=\omega_0$. Their coefficients, $\mathcal{S}$ and $\mathcal{A}$, depend on the field direction $\hat{\bm{h}}=\bm{H}/H$ and $\hat{\bm{s}}=\bm{s}/s$ as
\begin{subequations}
\label{eq:ST_FMR}
\begin{align} 
    \mathcal{S} &= (\hat{\bm{h}} \cdot \hat{\bm{s}}) \left\{ H_{\rm DL} [1-(\hat{\bm{h}} \cdot \hat{\bm{s}})^{2}] - H_{\rm Oe} (\hat{\bm{h}} \times \hat{\bm{y}} ) \cdot \hat{\bm{s}} \right\}, \label{eq:sym_S} \\
    \mathcal{A} &= \alpha'\mathcal{S} + (\hat{\bm{h}} \cdot \hat{\bm{s}}) \left\{ H_{\rm FL}[1-(\hat{\bm{h}} \cdot \hat{\bm{s}})^{2}] \right.  \nonumber \\
    &\qquad\qquad\qquad\qquad \left. + H_{\rm Oe}[ (\hat{\bm{h}}\times\hat{\bm{y}})\cdot(\hat{\bm{h}}\times\hat{\bm{s}})] \right\}, \label{eq:anti_sym}
\end{align}
\end{subequations}
where $\alpha' = (1 - g_{I}')(\alpha + g_{R}')/[(1 - g_{I}')^{2} - (\alpha + g_{R}')^{2}] \approx \alpha + g_{R}'\ll1$, so in Eq.~\eqref{eq:anti_sym} the $\alpha'\mathcal{S}$ term can be ignored compared with the rest. Should the Ørsted field be absent ($H_{\rm Oe}=0$), $\mathcal{S}$ and $\mathcal{A}$ will share the same angular dependence on the magnetic field, and their relative amplitude only relies on the ratio $H_{\rm DL}/H_{\rm FL}$, meaning that they exclusively indicate the DL and FL torques. This is consistent with the conventional SHE-based systems~\cite{Liu2011STFMR,macneill2017control}. However, when $H_{\rm Oe}$ is taken into account (meanwhile $\hat{\bm{s}}\neq\hat{\bm{y}}$ breaks the SHE relations), $\mathcal{S}$ and $\mathcal{A}$ exhibit quite different angular dependence on $\bm{H}$, distinct from their counterparts in the SHE limit.

\begin{figure}[t]
  \centering
  \includegraphics[width=\linewidth]{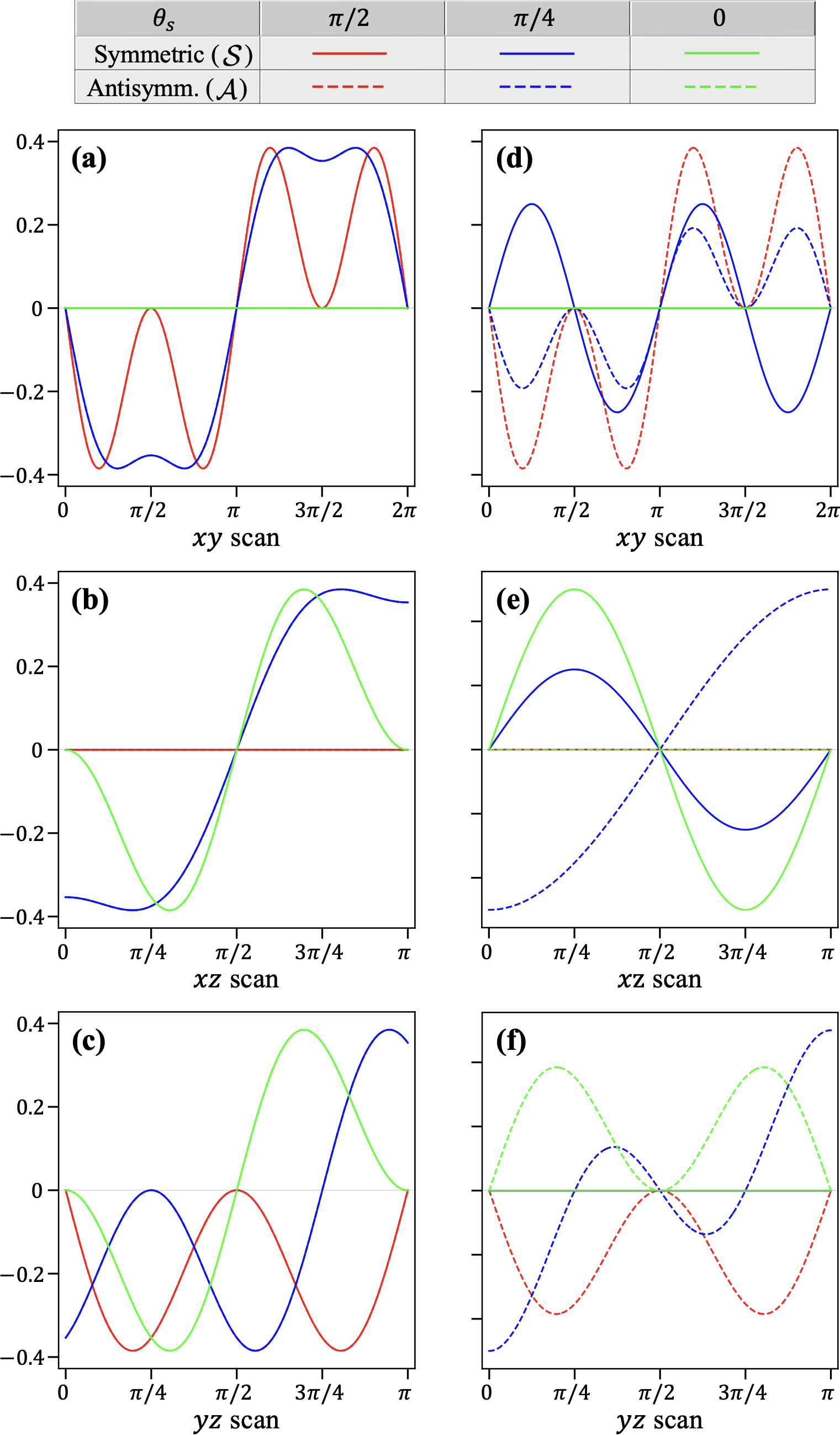}
  \caption{The coefficients of the symmetric and antisymmetric components of the rectification voltage $V_{\rm rec}/V_0$ with respect to $(H-\omega_0)$ plotted as functions of the angle of magnetic field undergoing the $xy$, $xz$ and $yz$ scans, for different values of $\theta_s$. (a)--(c): $H_{\rm Oe}=0$ while $H_{\rm DL}=H_{\rm FL}$ and the vertical axes scale in $H_{\rm DL}/\Delta$; (d)--(e): $H_{\rm DL}=H_{\rm FL}=0$ while the vertical axes scale in $H_{\rm Oe}/\Delta$.}
  \label{fig:ST_FMR_S_A}
\end{figure}

We consider WTe$_2$ again and let $\theta_s$ takes $\pi/2$ (the SHE limit), $\pi/4$ and $0$, respectively. To highlight the different contributions from the SOT and the Ørsted field to the angular dependence of the rectification signal on the magnetic field, we plot $\mathcal{S}$ and $\mathcal{A}$ in Fig.~\ref{fig:ST_FMR_S_A} with either the SOT or the Ørsted field (but not both) activated in each column of subfigures. The rows correspond to the $xy$, $xz$ and $yz$ scans, respectively. In all subfigures, the red curves mark the SHE limit, serving as a benchmark. The stark contrast between the red and other curves indicate the possibility to experimentally separate the \textit{t}-SHE from the ordinary SHE regarding the SOT generation. In Fig.~\ref{fig:ST_FMR_S_A}(a) and~(d), both $\mathcal{S}$ and $\mathcal{A}$ vanish for $\theta_{s} = 0$ (\textit{i.e.}, $\hat{\bm{s}}=\hat{\bm{z}}$) thanks to the $\hat{\bm{h}}\cdot \hat{\bm{s}}$ factor in Eq.~\eqref{eq:ST_FMR}, which is identically zero when $\hat{\bm{h}}$ rotates in the $xy$ plane. Similarly, if $\theta_{s} = \pi / 2$ (namely $\hat{\bm{s}} = \hat{\bm{y}}$), the factor $\hat{\bm{h}}\cdot \hat{\bm{s}}=0$ for the $xz$ scan, which explains the disappearing rectification signal in Fig.~\ref{fig:ST_FMR_S_A}(b) and (e). 

\begin{figure}[t]
  \centering
  \includegraphics[width=0.75\linewidth]{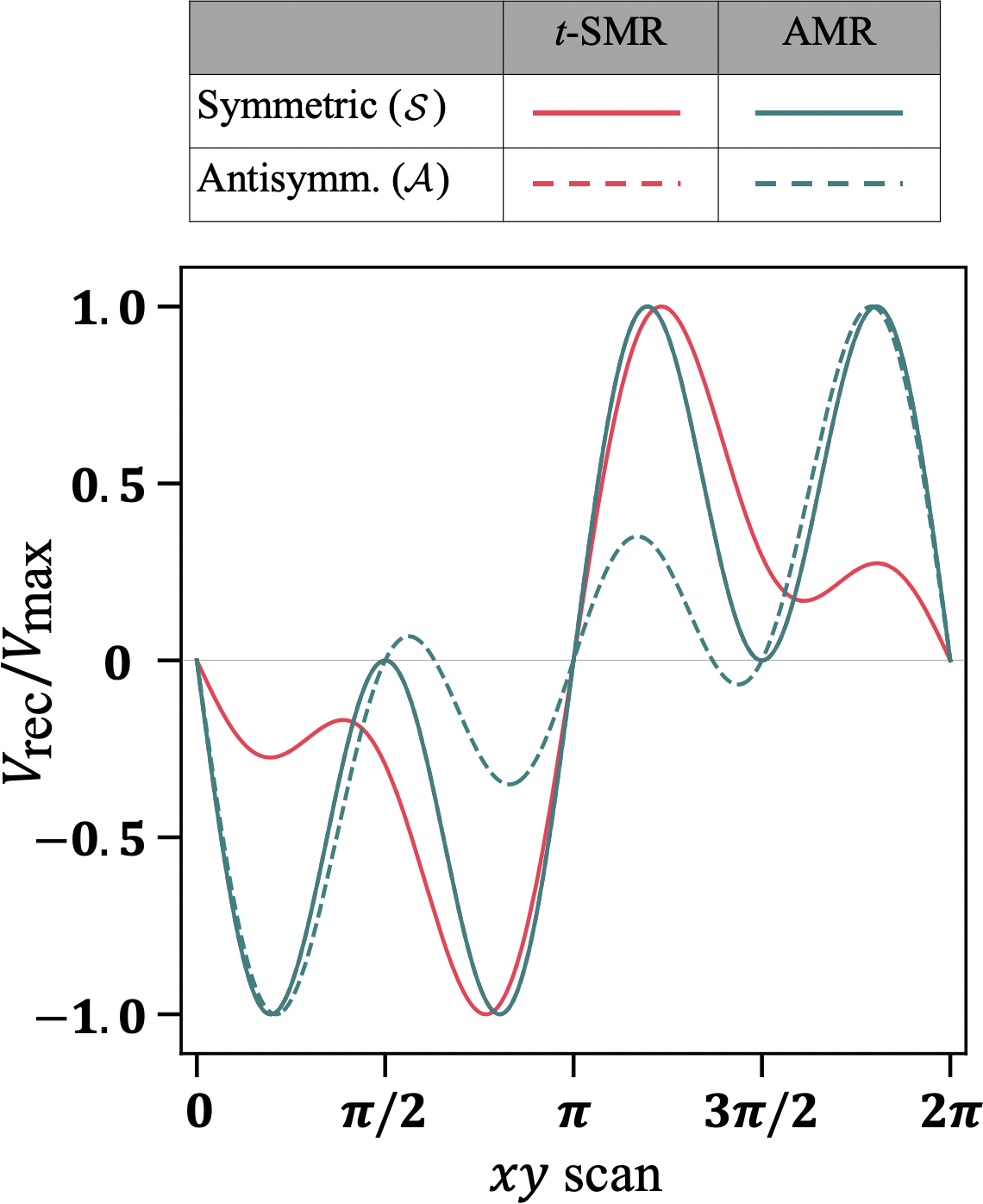}
  \caption{The symmetric and antisymmetric components of the rectification voltage with respect to $(H-\omega_0)$ plotted as functions of the angle of the magnetic field undergoing the $xy$ scan, where the red (green) curves correspond to the $t$-SMR (AMR) mechanism. $V_{\rm rec}$ is scaled by its maximum value $V_{\rm max}$, restricting the actual plot range within $\pm1$. The solid red ($\mathcal{S}$) and dashed red ($\mathcal{A}$) curves completely overlap.  
  }
  \label{fig:comp_to_exp}
\end{figure}

To further highlight the difference between the DC rectification originating from the $t$-SMR mechanism and that from the anisotropic MR widely reported in previous experiments, we compare the angular dependence on the magnetic field for these two distinct mechanisms. If one replaces the ferromagnetic insulator with a ferromagnetic metal~\cite{macneill2017control,macneill2017thickness}, however, the resulting harmonic signals will be dominated by the anisotropic MR rather than the $t$-SMR. Because AMR is an intrinsic property of ferromagnetic metals, its form is independent of the specific way of charge-to-spin conversion, in sharp contrast to the case of $t$-SMR. Such fundamental distinction necessarily reflects in the angular dependence of the rectification voltage. Using the experimentally characterized torque data from Ref.~\citenum{macneill2017control}, we plot $\mathcal{S}$ and $\mathcal{A}$ of the rectification voltage for both $t$-SMR and AMR as a function of the in-plane angle of the applied magnetic field, as shown in Fig.~\ref{fig:comp_to_exp}. In this plot, we set $\theta_s = 63.43^{\circ}$, $H_{\rm DL} / H_{\rm Oe} = 0.89$, and $H_{\rm FL} = 0$. While $\theta_s$ had not been directly reported in Ref.~\citenum{macneill2017control}, it could be indirectly estimated by the ratio between $\bm{H}_{\rm DL} \cdot \hat{\bm{y}}$ and $\bm{H}_{\rm DL} \cdot \hat{\bm{z}}$, which corresponds to $\tau_{S}$ and $\tau_{B}$ in Ref.~\citenum{macneill2017control}, respectively. Basing on the thickness dependence of the spin-orbit torques in Ref.~\citenum{macneill2017control}, we find that $\tau_{S}/\tau_{A} \approx 0.8$ and $\tau_{B} / \tau_{A} \approx 0.4$, which determines $\theta_{s} = \tan^{-1}(\tau_{S} / \tau_{B})$. The apparent discrepancy between our theoretical predictions and earlier measurements shown in Fig.~\ref{fig:comp_to_exp} underscores the necessity of our approach in guiding future experiments.

\textit{Second Harmonic.}---The second harmonic stemming from the $t$-SMR can be explored using a low-frequency AC drive and a sweeping magnetic field. Of particular importance concerning the characterization of SOT is the harmonic Hall signal $V_H^{2\omega}=\delta\bm{m}\cdot(\partial\overline{\rho}_{yx}/\partial\bm{m}_0)\mathcal{J}_x^{\rm(in)}/2$, which reveals critical information complementing the first harmonic and the rectification. Given a soft magnet coupled to a WTe$_2$ thin film as before, where $\hat{\bm{m}}$ is polarized by $\bm{H}$, we can solve the LLG Eq.~\eqref{eq:LLG_main} (see the SI) and obtain under the approximation $\tilde{g}_{I}\ll\tilde{g}_{R}$ that
\begin{subequations}
\label{eq:2nd_harmonic}
\begin{align} 
xy\mbox{ scan: } \frac{V_{H}^{2\omega}}{V_0} &= \frac{H_{\rm DL}}{2H} \sin 2\theta_{s} \sin^{2} \phi_{H} \nonumber\\
& + \frac{H_{\rm FL}}{H}\cos\phi_{H}(\sin^{2}\theta_{s}\cos2\phi_{H} + \cos^{2}\theta_{s}) \nonumber \\
& + \frac{H_{\rm Oe}}{H}\cos\phi_{H}\cos2\phi_{H}\sin\theta_{s}, \label{eq:2nd_harmonic_xy}\\
xz\mbox{ scan: } \frac{V_{H}^{2\omega}}{V_0} &= - \frac{H_{\rm DL}}{2H}\sin2\theta_{s}\cos^{2} \theta_{H} \nonumber\\
& + \frac{H_{\rm FL}}{H}\sin\theta_{H}(\sin^{2}\theta_{s} - \cos^{2}\theta_{s}\cos2\theta_{H}) \nonumber\\
& + \frac{H_{\rm Oe}}{H} \sin\theta_{H}\sin\theta_{s}, \label{eq:2nd_harmonic_xz} \\
yz\mbox{ scan: } \frac{V_{H}^{2\omega}}{V_0} &=\frac{H_{\rm DL}}{2H}\sin 2(\theta_{H} - \theta_{s}),
\label{eq:2nd_harmonic_yz}
\end{align}
\end{subequations}
where all notations follow the same definitions as in previous sections. If one considers a sizable $\tilde{g}_{I}$ (which could be comparable to or even larger than $\tilde{g}_{R}$), the above expressions will become very complicated.

\begin{figure}[t]
  \centering
  \includegraphics[width=\linewidth]{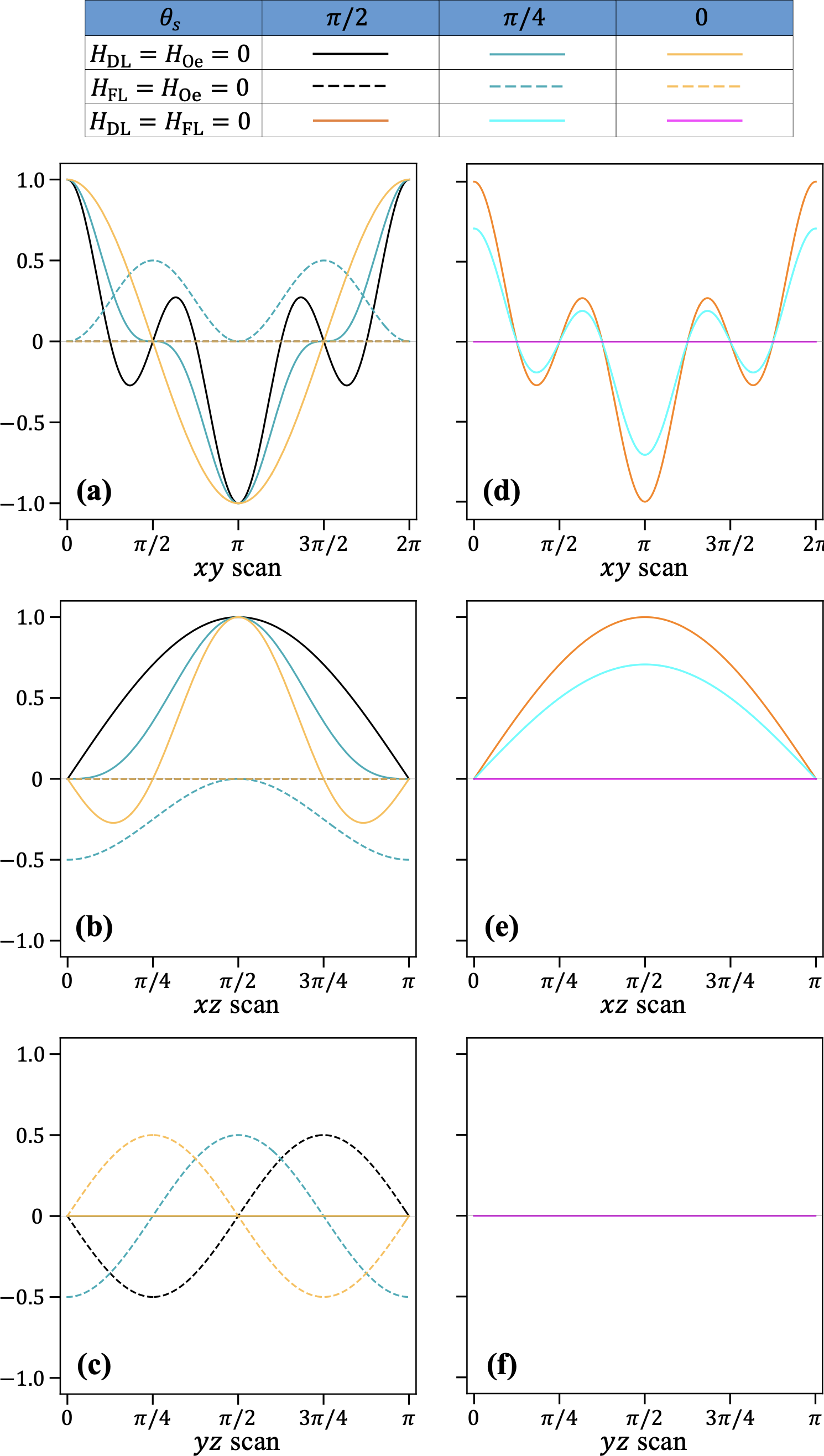}
  \caption{Second harmonic Hall voltage $V_{H}^{2\omega}/V_0$ as functions of the angle of magnetic field undergoing the $xy$, $xz$ and $yz$ scans for different values of $\theta_s$. (a)--(c): $H_{\rm Oe}$ vanishes and the solid (dashed) curves are plotted in the unit of $H_{\rm FL}/H$ ($H_{\rm DL}/H$) with $H_{\rm DL}=0$ ($H_{\rm FL}=0$), highlighting the contrast between the FL and DL torques. (d)--(e): $H_{\rm DL}$ and $H_{\rm FL}$ both vanish and the curves are plotted in the unit of $H_{\rm Oe}/H$.}
  \label{fig:2nd_harmonic_hall}
\end{figure}

When the low-symmetry axis of WTe$_2$ is collinear with the $x$ axis (\textit{i.e.}, the applied current direction), we know that $\hat{\bm{s}} = \text{cos}\theta_{s} \hat{\bm{z}} + \text{sin}\theta_{s} \hat{\bm{y}}$ and $\hat{\bm{s}}' = -\hat{\bm{x}}$. To separate and compare the contributions of $H_{\rm DL}$, $H_{\rm FL}$ and $H_{\rm Oe}$ to $V_H^{2\omega}$, we plot Eqs.~\eqref{eq:2nd_harmonic} in Fig.~\ref{fig:2nd_harmonic_hall} for different $\theta_s$, setting only one of $H_{\rm DL}$, $H_{\rm FL}$ and $H_{\rm Oe}$ to be non-zero in each individual curve. Similar to the case of first harmonic, the SHE limit (with $\theta_s=\pi/2$) serves as a benchmark in each subfigure. In particular, Fig.~\ref{fig:2nd_harmonic_hall}(c) and (f) suggests that only the DL torque contributes to $V_H^{2\omega}$ under the $yz$ scan, while the FL torque (and the Ørsted field) has no contribution, which is consistent with Eq.~\eqref{eq:2nd_harmonic_yz}.

In summary, we have established a comprehensive theoretical toolbox for experimental characterizations of the SOTs arising from the $t$-SHE, where the central quantity of interest is the $t$-SMR, marking a significant conceptual advance of the SHE and SMR widely used in existing studies. We anticipate our findings to timely inspire ongoing and future experiments.

\begin{acknowledgements}
The authors are grateful to Luqiao Liu, Simranjeet Singh, Andrew Kent, and Haoyu Liu for helpful discussions. This work was initially supported by the Air Force Office of Scientific Research under Grant No. FA9550-19-1-0307, and completed with the support of the National Science Foundation under Award No. DMR-2339315. All raw data corresponding to the findings in this paper is available upon reasonable requests.
\end{acknowledgements}

\newpage

\bibliography{reference}

\begin{thebibliography}{51}
\bibitem [{com()}]{commentAMR}%
  \BibitemOpen
  \href@noop {} {}\bibinfo {note} {In our model, the electric current flows in a non-SHE metal without shunting current; by contrast, in many recent studies involving WTe$_2$, the electric current flows largely within the ferromagnetic metal, and the measured resistance is mainly attributed to the anisotropic MR, where the underlying physical process is quite different.}\BibitemShut {Stop}%
\end{thebibliography}

\appendix*
\setcounter{equation}{0}
\section{Supporting Information}
\textit{Derivation of the t-SMR.}---The boundary conditions of the spin diffusion equation $\partial_{z}^{2} \mu_{j} = \mu_{j}/\lambda^{2}$ are
\begin{align}
 e\mathcal{J}_{zj}(z=0)&=g_R\epsilon_{jkl}\epsilon_{lab}\hat{m}_k\hat{m}_a\mu_b+g_I\epsilon_{jkl}\hat{m}_k\mu_l, \\
 \mathcal{J}_{zj}(z=t)&=0,
\end{align}
which was consistent with the ordinary SMR. Then by using Eqs.~\eqref{eq:current_potential}, we can solve the spin diffusion equation and obtain
\begin{align}
\label{eq:full_diffusion_sol}
 \mu_{j}(z) &= \mu_{j}^{0} \frac{\sinh [(2z - t)/2\lambda]}{\sinh (t /2\lambda)} - 2\lambda \frac{\cosh [(z-t)/\lambda]}{\sinh( t / \lambda) } \nonumber \\
&\qquad \times \left[\tilde{g}_{R}   \epsilon_{jkl}\epsilon_{lab}\hat{m}_k\hat{m}_a\mu_{b}^{0} + \tilde{g}_{I}      \epsilon_{jkl}\hat{m}_k\mu_{l}^{0})   \right],
\end{align}
where $\mu_{j}^{0} = 2et\eta\chi_{zjk}E_{k}$ with $E_{k}$ being the $k$ component of the applied electric field. Following a similar procedure as the ordinary SMR, we can derive from Eq.~\eqref{eq:full_diffusion_sol} the longitudinal and transverse components of the conductivity as
\begin{subequations}
\label{eq:pre_avg_sigma}
    \begin{align}
        &\frac{\sigma_{xx}}{\sigma} = 1 - \chi'_{xzk} \chi_{zkx} \frac{\cosh [(2z - t)/2\lambda]}{\cosh (t /2 \lambda)} + 2t \eta  \nonumber \\
        & \times \frac{\sinh [(z-t)/\lambda]}{\sinh (t /\lambda)} \left[ \tilde{g}_{R}( \chi_{zjx} \chi'_{xzk} \hat{m}_{j} \hat{m}_{k} -\chi_{zkx}\chi'_{xzk}  )  \right. \nonumber \\
        &\left. \qquad \qquad \qquad \qquad\qquad + \tilde{g}_{I} \epsilon_{kjl}\chi_{zlx}\chi'_{xzk} \hat{m}_{j}   \right] , \label{eq:pre_avg_sigma_xx} \\
        & \frac{\sigma_{xy}}{\sigma} = -\chi_{zkx}\chi'_{yzk}  \frac{\cosh [(2z - t)/2\lambda]}{\cosh (t /2 \lambda)} + 2t \eta \nonumber \\
        & \times \frac{\sinh [(z-t)/\lambda]}{\sinh (t /\lambda)} \left[ \tilde{g}_{R}( \chi_{zjx} \chi'_{yzk} \hat{m}_{j}\hat{m}_{k} -\chi_{zkx}\chi'_{yzk}  ) \right. \nonumber \\
        & \left. \qquad \qquad \qquad \qquad\qquad + \tilde{g}_{I}\epsilon_{kjl}\chi_{zlx}\chi'_{yzk} \hat{m}_{j} \right], \label{eq:pre_avg_sigma_xy}
    \end{align}
\end{subequations}
where $j$, $k$ and $l$ are summed over per Einstein's summation rule. Substituting $\chi_{ijk} = -\chi'_{kij}$ into Eqs.~\eqref{eq:pre_avg_sigma} and averaging the conductivity over the thickness direction, \textit{i.e.}, taking $(1/t)\int_0^t(\cdots)$, we obtain Eqs.~\eqref{eq:conductivities}.

\textit{Derivation of the rectification voltage}---The equilibrium direction of $\hat{\bm{m}}$ is simply $\hat{\bm{h}}$, and we define $\bm{m}_{\perp} = \hat{\bm{m}} - \hat{\bm{h}}$ such that $\bm{m}_{\perp} \cdot \hat{\bm{h}} = 0$ and $|\bm{m}_{\perp}| \ll 1$. Under the Fourier transformation, the LLG Eq.~\eqref{eq:LLG_main} becomes
\begin{align}
   &(1 - g_{I}') i \omega \tilde{\bm{m}}_{\perp} - [ H + (\alpha + g_{R}') i \omega] \hat{\bm{h}} \times \tilde{\bm{m}}_{\perp}  \nonumber \\
   &\qquad = \tilde{\bm{H}}_{\rm{DL}}-(\hat{\bm{h}} \cdot \tilde{\bm{H}}_{\rm{DL}}) \hat{\bm{h}} - \hat{\bm{h}} \times \tilde{\bm{H}}_{\rm{F}},
   \label{eq:LLG_m_perp}
\end{align}
where $\omega$ is the angular frequency of the driving current, $\tilde{\bm{m}}_{\perp}$, $\tilde{\bm{H}}_{\rm{DL}}$ and $\tilde{\bm{H}}_{\rm{F}}$ are the phasors of $\bm{m}_{\perp}$, $\bm{H}_{\rm{DL}}$ and $\bm{H}_{\rm{F}}$. By solving Eq.~\eqref{eq:LLG_m_perp}, we obtain 
\begin{align}
   \tilde{\bm{m}}_{\perp} =& \frac{1}{(1-g_{I}')^{2}\omega^2 - [H + (\alpha + g_{R}')i \omega]^{2}} \hat{\bm{h}} \times \nonumber \\
   & \{ (1 - g_{I}')i\omega  ( \hat{\bm{h}} \times \tilde{\bm{H}}_{\rm{DL}} + \tilde{\bm{H}}_{\rm{F}} ) \nonumber \\
   &\quad + [H + (\alpha + g_{R}')i \omega]
   ( \hat{\bm{h}} \times \tilde{\bm{H}}_{\rm{F}}  - \tilde{\bm{H}}_{\rm{DL}}) \}.
   \label{eq:m_perp}
\end{align}
According to Eq.~\eqref{eq:sigma_xx}, the longitudinal resistivity can be expressed as $\rho_{xx}(\tilde{\bm{m}}_{\perp}) = \rho_{0}(\hat{\bm{h}}\cdot\hat{\bm{s}})\rm{Re}(\tilde{\bm{m}}_{\perp} \cdot \hat{\bm{s}})$, where $\rho_{0}$ is an overall factor absorbing $\sigma$, $\tilde{g}_{R}$, $t$, $\eta$ and $|\bm{s}|$. Substituting Eq.~\eqref{eq:m_perp} into $\rho_{xx}(\tilde{\bm{m}}_{\perp})$ and using the near-resonance approximation $( H^{2} - \omega_{0}^{2})^{2} \approx 4\omega_{0}^{2} (H - \omega_{0})^{2}$, we arrive at Eqs.~\eqref{eq:ST_FMR_V} and \eqref{eq:ST_FMR}.

\textit{Derivation of 2nd harmonic Hall voltage}---In the low-frequency regime, the magnetizaion vector $\hat{\bm{m}}$ remains in quasi-equilibrium, so all time derivatives in Eq.~\eqref{eq:LLG_main} can be set to zero, yielding
\begin{equation}
\label{eq:f}
    \bm{f} = (\bm{H}  + \bm{H}_{\rm{F}}) \times \hat{\bm{m}} + \hat{\bm{m}} \times (\bm{H}_{\rm{DL}} \times \hat{\bm{m}}) = 0.
\end{equation}
In polar coordinates, $\hat{\bm{m}}=\sin \theta \cos \phi\hat{\bm{x}}+ \sin \theta \sin \phi\hat{\bm{y}}+ \cos \theta\hat{\bm{z}}$. To obtain the perturbed angles $\Delta \theta$ and $\Delta \phi$ in the presence of $\bm{H}_{\rm{F}}$, we first calculate $d \bm{f} / d \bm{H}_{\rm{F}}=0$:
\begin{align}
\label{eq:df_dHFL_orig}
    \left[\epsilon (\bm{H} + \bm{H}_{\rm{F}})  + \hat{\bm{m}} \otimes \bm{H}_{\rm{DL}}  +(\hat{\bm{m}} \cdot \bm{H}_{\rm{DL}}) \mathcal{I} \right] \frac{d \hat{\bm{m}}}{d \bm{H}_{\rm{F}}} \nonumber \\
    =- \hat{\bm{m}} \times \mathcal{I},
\end{align}
where $\epsilon$ the Levi-Civita tensor, $\mathcal{I}$ is the $3\times 3$ identity matrix, and $(\hat{\bm{m}} \times \mathcal{I})_{ij} = \epsilon_{iab} \hat{\bm{m}}_{a} \mathcal{I}_{bj}$. When Eq.~\eqref{eq:df_dHFL_orig} is evaluated in equilibrium that $\bm{H}_{\rm{F}} = \bm{H}_{\rm{DL}} = 0$, it can be rewritten as $\mathcal{M} (d \hat{\bm{m}} / d \bm{H}_{\rm{F}}) = \hat{\bm{m}} \times \mathcal{I}$ where the matrix $\mathcal{M}$ has components $\mathcal{M}_{ij} = -\epsilon_{ijk}H_k$. Then defining  the column vector $\bm{\theta} = (\theta, \phi)^{T}$ and using the chain rule $d\hat{\bm{m}}/d\bm{H}_{\rm{F}} = (d\hat{\bm{m}}/d\bm{\theta})( d\bm{\theta}/d\bm{H}_{\rm{F}})$, we have
\begin{align}
\label{eq:dtheta_dHFL}
    \frac{d \bm{\theta}}{d \bm{H}_{\rm{F}}}\!=\!\left[ \left(\mathcal{M}  \frac{d \hat{\bm{m}}}{d \bm{\theta}} \right)^{T}\left(\mathcal{M} \frac{d \hat{\bm{m}}}{d \bm{\theta}} \right) \right]^{-1}\!\!\left(\mathcal{M} \frac{d \hat{\bm{m}}}{d \bm{\theta}} \right)^{T}\!(\hat{\bm{m}} \times \mathcal{I}),
\end{align}
where
\begin{align}
\label{eq:dm_dtheta}
    \frac{d \hat{\bm{m}}}{d \bm{\theta}} =
    \begin{pmatrix}
    \cos \theta \cos \phi & -\sin \theta \sin \phi \\
    \cos \theta \sin \phi & \sin \theta \cos \phi \\
    -\sin \theta & 0
    \end{pmatrix}.
\end{align}
In equilibrium, $\hat{\bm{m}} = \hat{\bm{h}}$, so $\theta = \theta_{H}$ and $\phi = \phi_{H}$. We then obtain from Eqs.~\eqref{eq:dtheta_dHFL} and~\eqref{eq:dm_dtheta} that
\begin{align}
\label{eq:dtheta_dHFL_eq}
    \frac{d \bm{\theta}}{d \bm{H}_{\rm{F}}} = \frac{1}{H}
    \begin{pmatrix}
     \cos \theta_{H} \cos \phi_{H} & \cos \theta_{H} \sin \phi_{H} & -\sin \theta_{H} \\
     -\csc \theta_{H} \sin \phi_{H} & \csc \theta_{H} \cos \phi_{H} & 0 
    \end{pmatrix}.
\end{align}
Following the same procedure, we can also obtain 
\begin{align} \label{eq:dtheta_dHDL_eq}
    \frac{d \bm{\theta}}{d \bm{H}_{\rm{DL}}} = \frac{1}{H}
    \begin{pmatrix}
     \sin \phi_{H} & -\cos \phi_{H} & 0 \\
     \cot \theta_{H} \cos \phi_{H} & \cot \theta_{H} \sin \phi_{H} & -1 
    \end{pmatrix}.
\end{align}
Finally, the perturbed angles can be calculated as
\begin{align}
 \begin{pmatrix}
     \Delta\theta \\ \Delta\phi
 \end{pmatrix}
 = \frac{d\bm{\theta}}{d\bm{H}_{\rm{F}}} \bm{H}_{\rm{F}} + \frac{d\bm{\theta}}{d\bm{H}_{\rm{DL}}} \bm{H}_{\rm{DL}}.
\end{align}

After substitution of $\hat{\bm{s}} = \text{cos}\theta_{s} \hat{\bm{z}} + \text{sin}\theta_{s} \hat{\bm{y}}$ and $\hat{\bm{s}}' = -\hat{\bm{x}}$ into Eq.~\eqref{eq:sigma_xy}, the Hall resistivity $\rho_{xy}$ can be simplified to 
\begin{equation}
    \rho_{xy} =\rho_{0}( \sin \theta_{s}\hat{m}_{x}\hat{m}_{y} + \cos \theta_{s}\hat{m}_{x}\hat{m}_{z}),
    \label{eq:deltadelta}
\end{equation}
where $\hat{m}_{x}\hat{m}_{y}$ and $\hat{m}_{x} \hat{m}_{z}$ are
\begin{subequations}
\label{eq:mxmymz}
\begin{align}
    &\hat{m}_{x}\hat{m}_{y} = \frac{1}{2}\sin^{2}\theta \sin 2\phi \approx \frac{1}{2} (\sin^{2}\theta_{H} \sin 2\phi_{H} \nonumber \\
    &  + 2\sin^{2} \theta_{H} \cos 2\phi_{H} \Delta \phi + \sin 2\theta_{H} \sin 2\phi_{H} \Delta \theta), \label{eq:mxmy} \\
    & \hat{m}_{x} \hat{m}_{z} = \frac{1}{2} \sin 2\theta \cos \phi \approx \frac{1}{2} ( \sin 2\theta_{H} \cos \phi_{H}  \nonumber \\
    & - \sin 2\theta_{H} \sin \phi_{H} \Delta \phi + 2 \cos 2\theta_{H} \cos \phi_{H} \Delta \theta  ). \label{eq:mxmz}
\end{align}
\end{subequations}
Plugging $\Delta \theta$ and $\Delta \phi$ obtained from Eq.~\eqref{eq:deltadelta} into Eqs.~\eqref{eq:mxmymz}, we can justify Eqs.~\eqref{eq:2nd_harmonic}.

\end{document}